# Title: How Jupiter's Unusual Magnetospheric Topology Structures Its Aurora

**Authors:** Binzheng Zhang[1,2]*, Peter A. Delamere[3], Zhonghua Yao[4]*, Bertrand Bonfond[5], D. Lin[2], Kareem A. Sorathia[6], Oliver J. Brambles[7], William Lotko[2,8], Jeff S. Garretson[6], Viacheslav G. Merkin[6], Denis Grodent[5], William R. Dunn[9] and John G. Lyon[6,10]

**Affiliations:**

[1] Department of Earth Sciences, the University of Hong Kong, Hong Kong SAR, China.

[2] High Altitude Observatory, National Center for Atmospheric Research, Boulder, CO, USA.

[3] Geophysical Institute, University of Alaska Fairbanks, Fairbanks AK, USA.

[4] Key Laboratory of Earth and Planetary Physics, Institute of Geology and Geophysics, Chinese Academy of Sciences, Beijing, China

[5] LPAP, Space sciences, Technologies and Astrophysics Research (STAR), Institute Université de Liége (ULiége), Belgium

[6] Applied Physics Laboratory, Johns Hopkins University, Laural MD, USA

[7] Oliver J. Brambles Ph.D., Preston, UK

[8] Thayer School of Engineering, Dartmouth College, Hanover NH, USA

[9] Mullard Space Science Laboratory, University College London, Dorking, UK

[10] Gamera Consulting, Hanover NH, USA

*Correspondence to: binzh@hku.hk and zhonghua.yao@uliege.be

**Abstract:** Jupiter's bright persistent polar aurora and Earth's dark polar region indicate that the planets' magnetospheric topologies are very different. High-resolution global simulations show that the reconnection rate at the interface between the interplanetary and jovian magnetic fields is too slow to generate a magnetically open, Earth-like polar cap on the timescale of planetary rotation, resulting in only a small crescent-shaped region of magnetic flux interconnected with the interplanetary magnetic field. Most of the jovian polar cap is threaded by helical magnetic flux that closes within the planetary interior, extends into the outer magnetosphere and piles-up near its dawnside flank where fast differential plasma rotation pulls the field lines sunward. This unusual magnetic topology provides new insights into Jupiter's distinctive auroral morphology.

**One Sentence Summary:** Jupiter's slow dayside magnetic merging and fast rotation produce an unusual magnetospheric topology and distinctive auroral morphology.

**Main Text:**

Impressive auroral displays are seen at every magnetized planet with an atmosphere, but not all aurora are created equal. Earth's aurora is episodic with a usually well-defined oval of bright ultraviolet (UV) luminosity surrounding a dark polar region above 70-75° magnetic latitude (MLAT). Jupiter also has an auroral oval encircling the magnetic poles (Figure 1), but unlike Earth's, it is persistent, and its polar cap—the region poleward of the auroral oval—contains bright, dynamic aurora that account for about half of the emitted power (*1, 2*). Jupiter's polar aurora is often grouped into at least three structures, including a "swirl region" in the center, peppered with dim, intermittent and chaotic bursts of emissions, an "active region" of flares, bright spots and arc-like structures on the duskside (*1, 3, 4*), and a "dark region" located on the dawnside (*5, 6*).

Auroras are produced by energetic charged particles that excite atomic emissions when precipitating into the atmosphere. Most of these particles are repelled and trapped in space by the mirror force of the planetary magnetic field, except the few that are forced or scattered into the so-called loss cone and precipitate (*7*). Observed morphological differences in aurora are thus signatures of the different magnetic topologies that define the planetary space environment (the magnetosphere) and the different processes that enable auroral precipitation (*8*). The connection between Earth's aurora and its magnetospheric topology has been explored extensively and is reasonably well-understood (*1,2*). The jury is still out on the magnetic structure of Jupiter's magnetosphere and what exactly its aurora is telling us about its topology.

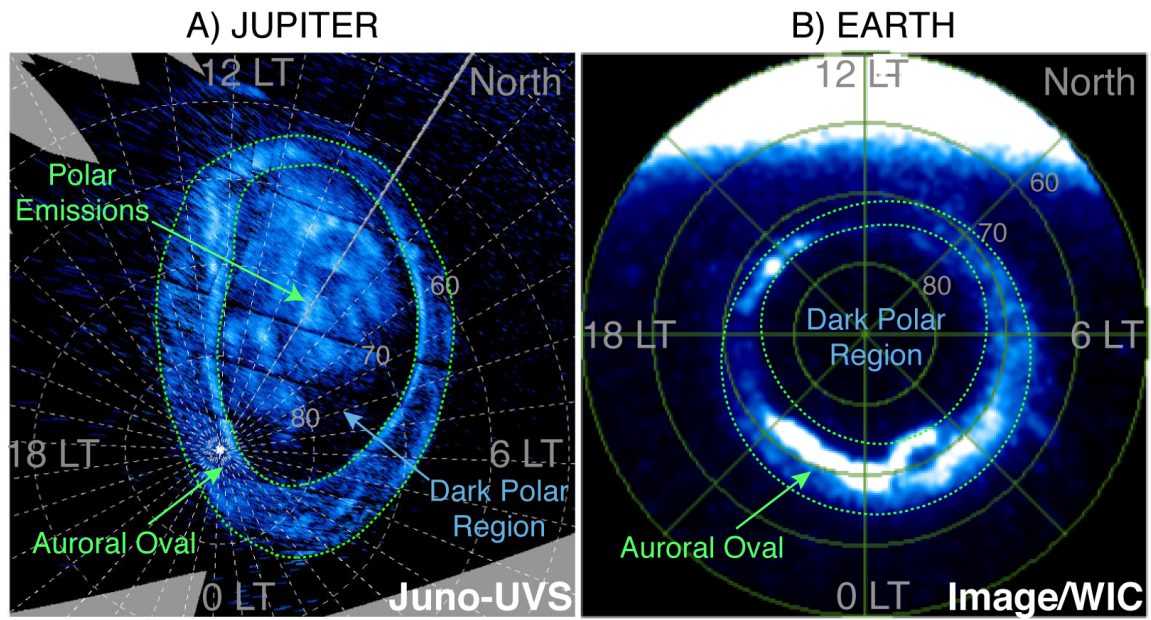

**Fig. 1.** Polar projections of Northern UV aurora at Jupiter (A) and Earth (B) vs. latitude and local time. Juno-UVS image was acquired on 19/05/2017 at 04:21:56. WIC image was acquired on 14/01/2001 at 05:00:55UT.

The magnetic flux threading Earth's polar cap is typically "open" and interconnected with the interplanetary magnetic field (IMF), a consequence of dayside magnetic reconnection wherein dissipative merging between the IMF and geomagnetic field breaks the frozen-in condition of ideal magnetohydrodynamics (*9*). The polar cap is dark because electrons precipitating into the atmosphere from the extremely low-density plasma populating open field lines have insufficient energy flux to excite intense auroral emissions (*10*). Earth's auroral oval occurs on closed magnetic field lines, meaning the magnetic flux in geospace traces poloidal paths between hemispheres (*11*). The auroral oval is the dominant source of the terrestrial UV auroral emissions, with > 70% of the emitted power typically coming from the so-called diffuse aurora and the remainder in the more structured and bright discrete and wave-induced aurora (*12*).

The electron precipitation responsible for Jupiter's auroral oval resembles the discrete and wave-induced aurora at Earth. The causative downward electron flux carries upward-directed magnetic field-aligned currents (FAC), largely generated by the breakdown of plasma corotation on closed magnetic field lines extending (beyond ~20 $R_J$) (*13, 14*), a consequence of the Vasyliunas cycle driving mid- and low-latitude plasma circulation (*15*). In analogy with Earth, open field lines are thought to thread most of the region poleward of Jupiter's auroral oval, but this open-flux model is difficult to reconcile with observed precipitation of energetic electrons (*16*) over the polar region, along with ions released by the volcanic moon Io (*17-20*).

Thus, Jupiter's bright polar aurora presents quandaries. If it threads open magnetic field lines, why is the precipitating particle energy flux along Jupiter's open field lines so much greater than expected, and how does heavy-ion precipitation access open field lines? A polar cap threaded in part by closed magnetic field lines eliminates these quandaries, but raises questions about the origins of the polar magnetic topology, especially in light of observed reconnection signatures on the dayside and dawnside magnetopause (magnetospheric interface with the IMF) resembling those at Earth (*17, 18*). Theoretical studies suggest that the rate of large-scale reconnection at Jupiter may not be fast enough to produce a fully open polar cap (*19, 20*) with very few polar field lines interconnected with the IMF (*21, 22*). But is it not clear how or if the polar-region open and closed magnetic field lines are linked and distributed in the magnetosphere?

These questions are not easily addressed with the limited in-situ observations available for Jupiter. Physics-based model simulations offer an interpretive framework for the observations, particularly the recent in situ measurements from the Juno spacecraft. We investigated the magnetic topology of Jupiter's polar cap using a newly developed global magnetohydrodynamic (MHD) model of the jovian magnetosphere, including its interactions with the interplanetary medium, the Io plasma torus, and ionosphere-magnetosphere coupling (*23-25*). The new results reported here offer a testable model of Jupiter's polar magnetic topology and its magnetic connectivity to the jovian outer magnetosphere and interplanetary medium.

We specified time-stationary, idealized upstream conditions corresponding to a typical "non-compressed" magnetosphere, formed by a Mach 10 solar wind (SW) with number density 0.2 cm$^{-3}$, speed 400 km/s, dynamic pressure 0.03 nPa and an east-west IMF component 0.5 nT (*26*). The heavy-ion mass loading from Io is set to 1000 kg/s based on empirical estimates (*27*). The simulation was run for 230 hours (23 jovian days) with results reported here derived from the last two planetary spins. We determined the amount of open magnetic flux in the simulated jovian magnetosphere, the connection and linkages of magnetic field lines emanating from the polar and auroral regions, and implications for jovian auroral features.

The reconnection potential $\mathcal{E}_{reconn}$ and open magnetic flux $\Phi$ are related through the Faraday's law:

$$\frac{\partial \Phi}{\partial t} = -\int_{MP} \boldsymbol{E} \cdot \boldsymbol{dl} = \mathcal{E}_{reconn}.$$

$\boldsymbol{E}$ is the electric field along the magnetopause (MP) separatrix—the locus of points separating open and closed magnetic field lines (Figure 2A) (28, 29). Integration of the electric potential $\boldsymbol{E} \cdot \boldsymbol{dl}$ projected along the magnetopause separatrix (black curve in Figure 2B) yields a reconnection potential of $\mathcal{E}_{reconn} \approx 508$ kV. This reconnection potential is within the range of estimates based on interplanetary measurements near the Jupiter (30).

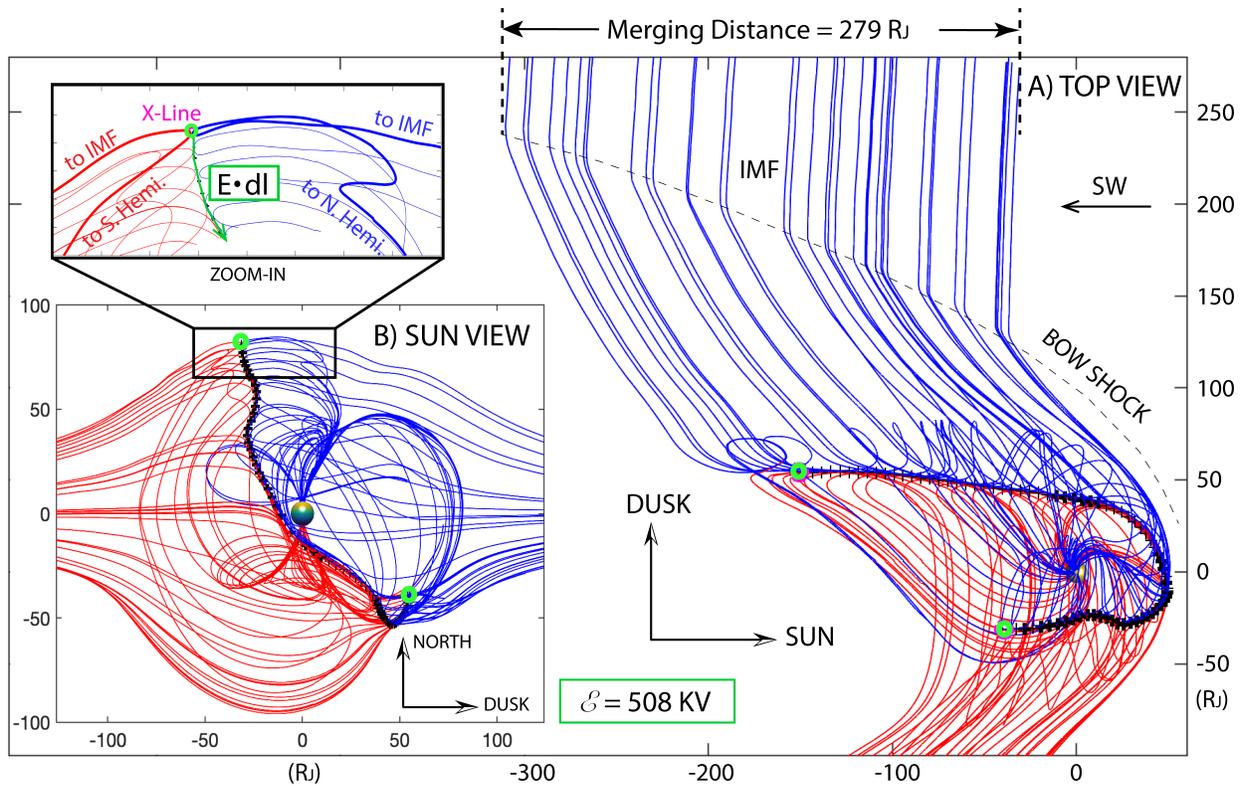

**Fig. 2.** Jovian magnetic field lines connected to the IMF after undergoing reconnection at the magnetopause. Black curves (magnetic separatrix) separate topological classes: open, closed and IMF. View from: Northern hemisphere down onto the pole (A), Sun (B), with zoom-in example of a reconnection site. Blue/red field lines emerge from the northern/southern hemisphere. The green circles locate the termination of the magnetopause reconnection separatrix.

In quasi-steady state, the transit time (Δt) for SW advection of a newly reconnected and open field line at the dayside MP to the nightside, where it undergoes reconnection again to become a closed field line, is determined from the simulated spatial extent of open flux in the SW (blue lines in Figure 2) divided by the SW speed: Δt = 287.7 R$_J$/400 km/s ≈ 48100 s (13.3 h). This transit time with the above-calculated $\mathcal{E}_{reconn}$ in Faraday's law then determines the open flux created by magnetopause reconnection as

$$\Delta\Phi \approx \Delta t \cdot \mathcal{E}_{reconn} \approx 24.4 \text{ GWb}.$$

This open flux is approximately 9% of the total dipole magnetic flux (259 GWb) threading the simulated jovian PC, taken to be the area in Fig. 3a above ≈ 82° magnetic latitude (MLAT). Flux-equivalence mappings of the low-altitude footpoints of magnetic fields measured in the jovian magnetosphere (*31*) suggest an 11° symmetric-circle equivalent of the observationally ill-defined, asymmetric PC. The mapped equivalent flux within the PC, assumed to be fully open in (*31*), is estimated at 700-730 GWb and is about 50% larger than the dipole equivalent flux of 488 GWb for a PC within 11° of the pole. The generation of such large open flux, given the simulation value for $\mathcal{E}_{reconn}$, requires a magnetopause merging distance of order 4 AU, wherein the field lines most likely become Alfvénically disconnected from Jupiter. Therefore, the spatial distribution of Jupiter's open PC flux must be very different from that of Earth's PC.

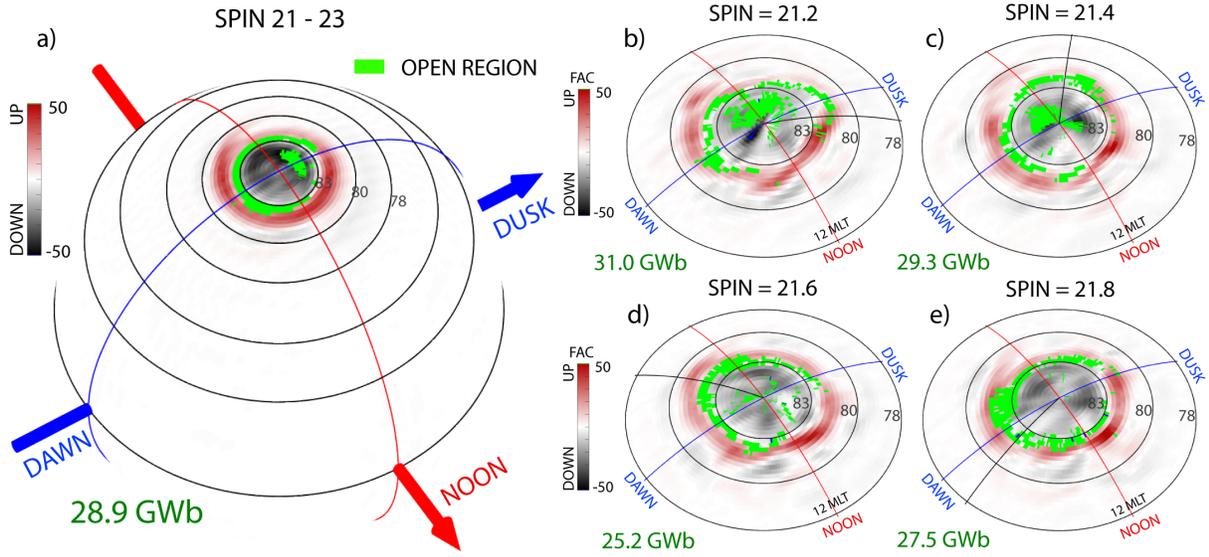

**Fig. 3.** Distributions of PC open flux and FAC density. A: Distributions averaged over simulation days 21- 23. Instantaneous distributions at diurnal time 21.2 (B), 21.4 (C), 21.6 (D) and 21.8 (E). The open flux region is shown in green. Upward (downward) FACs are shown in red (grey). Projections are on the 6 $R_J$ spherical surface vs. MLAT and LT with total open flux (in GWb) given at lower left.

Figure 3A shows the spatial distribution of open flux and FACs derived from the magnetic field averaged over 2 jovian days and traced to the low-altitude boundary of the simulation. Instead of forming a single circular-shaped open PC as at Earth, the simulated average open flux at Jupiter threads two disjoint polar regions: 1) a crescent-shaped region spanning ≈ 82-83° magnetic latitude (MLAT) and extending from dusk to noon in local time (LT) between regions of upward and downward FAC and 2) a duskside patch region above 85° MLAT. The crescent-shaped open flux is magnetically connected to the dawnside IMF (Figure 2A) with an average magnetic flux of 23.4 GWb in excellent agreement with the above estimate from Faraday's law. The average spatial distribution of the crescent-shaped open flux is similar to the crescent-shaped polar region devoid of auroral emissions in Hubble Space Telescope (HST) images (*32*). The simulated crescent exhibits less MLAT distortion than the void in HST images, probably due to the use of an axisymmetric dipole magnetic field; however, the simulated and observed crescents are both more or less fixed in local time, in contrast with other polar features (*33*). The high-latitude patch region above 85° MLAT contains about 19% (5.5 GWb) of the total open flux. If the simulated PC in Figure 3A is taken to be the area above 82° MLAT, then 89% of the PC flux is closed in the simulation.

Both the crescent and patch regions of the jovian open flux exhibit dynamic variations. Figure 3B-E show snapshots of the distributions of open flux together with FACs derived at different phases of planetary spin 21. The open flux of the instantaneous crescent-region near 82-83° MLAT varies from 18.6-25.2 GWb. The patch region is more intermittent with a highly variable spatial distribution, e.g., its open flux is 3.7 - 11.0 GWb during the first half of spin 21 and almost disappears during the second half of spin 21 (0.3 - 3.7 GWb). What exactly controls the patch region deserves further investigation. It is likely generated via complicated interactions between magnetotail reconnection and ionospheric electrodynamics, regulated by the strength of the dipole, angular speed of planetary rotation, orientation of the upstream IMF and spatial gradients in ionospheric conductivity (*34*).

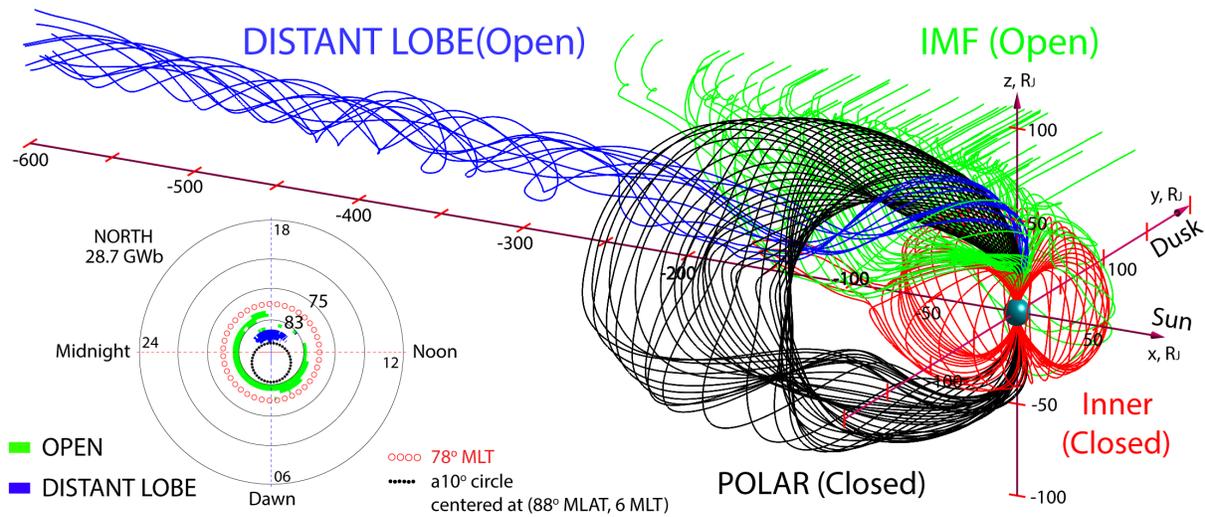

**Fig. 4.** Different topological classes of jovian magnetic field lines averaged over simulation days 21-23. Illustrative field lines emerge from four low-altitude sets in the northern hemisphere (lower left panel). Red: 78° MLAT. Green: Crescent region. Blue: Patch region. Black: Closed Polar Cap.

Figure 4 shows the two-day average topology of simulated jovian magnetic field lines traced from four different sets of footpoints at the northern low-altitude simulation boundary. The green open field lines have footpoints in the crescent region and are connected to the IMF. They are created by magnetopause reconnection. The blue field lines originate from the duskside patch region and map to the distant tail beyond 600 $R_J$. The downward FACs flowing along these helical open field lines (winding counterclockwise when looking upward) are generated by differential rotation of the polar ionosphere (*35, 36*). Their long extent is consistent with observations of Jupiter's very elongated magnetotail (*37*). The red closed field lines have footpoints at 80° MLAT. They map outward to 35-40 $R_J$ in the middle magnetosphere and return to the southern boundary (ionosphere). The black field lines emerging from Jupiter's polar region illustrate one of the more unusual features of jovian magnetic topology. These helical, closed field lines connect the two polar regions through the dawnside outer magnetosphere and have no counterpart in the terrestrial magnetosphere. This topological feature occurs because the time scale for reconnection to retain open flux, estimated above as 48100 s, is too slow to generate a complete open PC in a single jovian rotation of period 36540 s. Thus the black polar field lines in Fig. 4 cannot access the IMF at the magnetopause and remain closed. These magnetic flux pile-up on the dawnside flank, are bent sunward by Jupiter's differentially rotating plasma and develop a low-latitude boundary layer through viscous stresses at the magnetopause boundary (*22*). Their geodesic curvature is produced by field-perpendicular currents that flow toward the planet, balance the inertia of differential plasma rotation via a tailward MHD Lorentz force and are diverted into downward PC field-aligned currents (Fig. 3), in a manner similar to what occurs in Earth's low-latitude boundary layer (*38*). This giant volume of trapped jovian magnetospheric plasma connects the dynamics of the two polar hemispheres. Its variability stimulates Alfvénic perturbations, fieldline resonant oscillations and associated particle acceleration to power polar aurora. Observed 2-3 minute oscillations of jovian polar auroral emissions (*3*) arise naturally on these closed field lines, which should also support hemispherically conjugate aurora.

Jupiter's unusual magnetospheric topology requires care when interpreting the sources of the planet's polar aurora. For example, some or all of the dayside "active region" (*39*) maps along closed polar field lines (Fig. 4) into the dawn-to-midnight equatorial magnetosphere. Auroral activity in this region might then be attributable to flankside boundary layer and/or nightside magnetotail dynamics rather than dayside reconnection, as is typical at Earth. The new insights into jovian magnetic field topology reported here are not only important in the interpretation of jovian auroral processes, but they also applicable to other rapid rotating planets such as Saturn, Uranus and Neptune.

**Acknowledgments:**

**Funding**: B. Z is supported by the RGC Early Career Scheme (27302018) and the General Research Fund (17300719). D.G. is supported by the PRODEX program managed by ESA in collaboration with the Belgian Federal Science Policy Office. Computing resources were provided by NCAR CISL Grants UJHB0015.

**Data and materials availability:** Simulation codes and data are available from B. Zhang.

# Supplementary Materials for

# How Jupiter's Unusual Magnetospheric Topology Structures Its Aurora

Binzheng Zhang, Peter A. Delamere, Zhonghua Yao, Bertrand Bonfond, Dong Lin, Kareem A. Sorathia, Oliver J. Brambles, William Lotko, Jeff S. Garretson, Viacheslav G. Merkin, Denis Grodent, William R. Dunn and John G. Lyon

correspondence to: binzh@hku.hk and zhonghua.yao@uliege.be

**This PDF file includes:**

Methods
Figs. S1 to S4

## Methods

### *A. The global Jovian magnetosphere model*

#### *A.1 The GAMERA code and the computational grid*

We use the Grid Agnostic MHD for Extended Research Applications (GAMERA) global model (*24*) to simulate the interaction of the solar wind and (SW) interplanetary magnetic field (IMF) with the Jovian space environment – the volume of space where the Jovian magnetic field dominates the IMF. The model is based on equations of multi-fluid MHD (*40, 41*), including: a) a dipole magnetic field embedded at the center of the Jupiter to represent the planet's intrinsic magnetic field; b) supersonic upwind conditions representing the SW and embedded IMF powering the interaction; c) low-altitude boundary conditions representing the closure of magnetospheric field-aligned currents in the Jovian ionosphere; d) rotation of the planet at the low-altitude boundary imposed through an electrostatic potential; e) a plasma source representing heavy ion mass loading from the Io plasma torus, and f) numerical resistivity that enables magnetic reconnection in ideal MHD.

The computational volume of the simulation is a stretched sphere of length 1200 $R_J$ along the *x*-axis of the solar magnetic (SM) coordinate system of the Jupiter and 400 $R_J$ in the directions perpendicular to the *x*-axis. A spherical volume of radius 6 $R_J$ is cut out inside the distorted spherical computational domain. The sphere is centered on Jupiter, 100 $R_J$ downstream from the upwind surface where the SW/IMF conditions are imposed. Figure S1 shows the three-dimensional view of the computational grid.

The primary computational technique is explicit, finite volume MHD. The code uses Adams-Bashforth time marching with an upwind, seventh-order spatial reconstruction. In addition, non-linear numerical switches based on the Partial Donor Method are used to maintain the total variation diminishing (TVD) property. A staggered grid is used to achieve zero numerical divergence of the magnetic field while fulfilling this TVD condition for the system of equations (*42*). The finite volume technique allows the code to complete its calculation on a non-orthogonal numerical grid with cells adapted to the configuration of the Jovian magnetosphere, e.g., cells that are smaller in the inner magnetosphere and across the nominal magnetopause than parallel to it. The computational grid used for the runs in this paper has 256×256×256 cells. The grid is spherical polar near the low-altitude boundary with axis of symmetry along the SM x axis and with indices (i×j×k) corresponding approximately to the spherical (r×θ×ϕ) coordinates (radial×meridional× azimuthal) with horizontal resolution mapped to the ionosphere of approximately 0.3°×0.15° in MLT and MLAT.

A numerical resistivity enables magnetic reconnection in the code when and where magnetic gradients reach the grid scale, which is approximately 0.16 $R_J$ near the x-line on the dayside magnetopause for the 256×256×256 resolution computational grid. The resulting reconnection electric field is typically of order $0.1 v_A B_{in}$, in terms of the Alfvén speed $v_A$ and magnetic field $B_{in}$ in the reconnection inflow region. This practically universal scaling indicates that the reconnection rate is not caused by changes in numerical dissipation and that the numerical resistivity is simulating large-scale aspects of reconnection appropriately (*28*). It has also been shown that the global rate of dayside reconnection in the magnetosphere is controlled by upstream conditions in the global simulation model rather than the grid resolution of the simulation (*43*).

*A.2 Implementation of Mass Loading from the Io*

The GAMERA code uses a symmetric, eight-cell stencil to reconstruct the numerical fluxes at cell interfaces, which means that the flux through the low-altitude interface depends on the four boundary (ghost) cells as well as the first four active computational cells in the radial direction. Rather than modifying the plasma variables in the ghost cells, the mass-loading module directly manipulates the numerical fluxes at the interface of the inner boundary in order to preserve conservation of mass. This interface flux implementation allows the code to inject the exact amount of heavy ($O^+$) ions as specified into the active simulation domain. The hard-wall boundary condition keeps plasma populations already in the active domain from leaking back to the ghost zones through the inner boundary (*44*).

The Io plasma torus for mass loading at the inner boundary interface is chosen to be a spatially uniform band centered in the equatorial plane, with $\pm 0.5$ $R_J$ extent in the z-direction of the SM coordinates. Figure S2A shows the spatial distributions of the $O^+$ number density at t = 0.1 (spin), which gives a total rate of approximately 1000 kg/s as suggested by empirical analysis (*27*). The radial momentum flux for $O^+$ is set to zero, while the tangential momentum fluxes of $O^+$ are set based on the corotation speed. The energy flux in the mass loading module is then calculated based on a fixed $O^+$ temperature of 50 eV. Figure S2B- S2C show the corresponding spatial distributions of $O^+$ number density after the mass loading module switched on for 11 and 22 planetary spins, respectively. It is clear that the spatial distribution of the introduced plasma population exhibits a "torus-like" structure during the whole simulation due to the high resolving power of the numerical schemes used in GAMERA, without significant numerical spreading of the torus plasma in the z-direction.

*A.3 Implementation of Magnetosphere-Ionosphere (MI) Coupling and Co-rotation*

M-I coupling is implemented by combining Ohm's law with current continuity and the electrostatic approximation in the ionosphere, to obtain the following two-dimensional elliptic equation for the ionospheric electric potential $\Phi_i$, given the field-aligned current $J_{\|i}$ at the top of the ionospheric conducting layer (assumed to be $\approx 1$ $R_J$ and the height-integrated conductance tensor $\Sigma$):

$$\nabla \cdot \Sigma \cdot \nabla \Phi_i = J_{\|i} \cos\alpha \qquad (1)$$

The dip factor $\cos\alpha$ is $\mathbf{b} \cdot \mathbf{r_0}$, where $\mathbf{b}$ is a unit vector pointing along the dipole magnetic field at the top of the ionospheric conducting layer and $\mathbf{r_0}$ is the radial unit vector in spherical polar coordinates. Field-aligned current $J_{\|i}$ is computed near the low-altitude computational boundary at approximately 6.1 $R_J$ Jovicentric in the magnetosphere and is then mapped along dipole field lines assuming $J_\|/B$ = const to the top of the height-integrated ionospheric conducting layer to give $J_{\|i}$. The height-integrated substrate is located at 1.01 $R_J$ Jovicentric where equation (1) is solved. The ionospheric electric potential $\Phi_i$ obtained from equation (1) is mapped along dipole field lines to the inner computational boundary (6 $R_J$ Jovicentric) assuming the magnetic field lines are equal-potential. The electric field at the boundary is calculated as $\mathbf{E}_\perp = -\nabla_\perp \Phi_i$ which serves as a part of the low-altitude boundary condition for the MHD solver. The ionospheric potential equation is solved with a newly developed code, dubbed REMIX, largely based on the legacy Magnetosphere-Ionosphere Coupler/Solver (MIX) program (*45*). As a first step, we neglect the gradient of the Hall conductance and use $\Sigma_P = 0.1$ S as an approximation for the Jovian ionosphere (*46*).

After solving the convective potential $\Phi_i$, the implementation of corotating magnetospheric plasmas and flux tubes is equivalent to adding a tangential electric field component at the inner boundary of the simulation domain induced by the rotation of the Jupiter:

$$E_{cr} = -(\mathbf{\Omega_J} \times \mathbf{r}) \times \mathbf{B}, \qquad (2)$$

where $\Omega_J = 1.76 \times 10^{-4}$ rad·s$^{-1}$ is the angular speed of Jupiter's rotation, corresponding to a 9.9 hours period. At Jupiter's ionospheric reference altitude ($\approx R_J$) with equatorial magnetic field strength $B_J$, the corotation of the Jovian magnetosphere is implemented by imposing a time-stationary corotation potential $\Phi_{coro}$ at the ionospheric boundary given by:

$$\Phi_{coro} = -\Omega_J B_J R_J^2 \sin(\lambda), \qquad (3)$$

where $\lambda$ is the magnetic co-latitude, $B_J = 4.27$ G, $R_J = 69911$ km. This corotation potential is combined with the electrostatic potential $\Phi_i$ solved through Equation (1) in the REMIX module and dipole mapped to the inner boundary of the MHD domain:

$$\Phi_{total} = \Phi_i + \Phi_{coro}. \qquad (4)$$

Then the electric field driving the corotation at the inner boundary is calculated as $E = -\nabla \Phi$ total. Figure S3 shows the spatial distribution of the time-stationary $\Phi_{coro}$ and a snapshot of $\Phi_i$ at t = 203.5 hour at the interface of the inner boundary (6 $R_J$). Note that the peak value of $\Phi_{coro}$ is -70967 kV, while the peak value of $\Phi_i$ is only -643 kV, suggesting that the ionospheric "convection" potential is in general much smaller than the corotation potential, which is consistent with previous estimations (*47*). The value of $\Phi_i$ is consistent with the magnetopause reconnection potential, which is approximately 586 kV calculated based on the instantaneous electric and magnetic fields. Above 83 degrees MLAT, the peak corotation potential is approximately 6200 kV, which is still about one order of magnitude greater than the convection potential. The comparison between the corotation potential $\Phi_{coro}$ and the ionospheric convection potential $\Phi_i$ shows that the Jovian magnetosphere is largely driven by the rotation of the planet, and the solar wind-magnetosphere interaction may be a secondary effect when driven by nominal upstream conditions.

## B. Determination of the reconnection separatrix

### B.1 The hemispheric-marching method

Magnetic reconnection occurs on the magnetopause where the incoming IMF have a sufficient shear angle relative to the intrinsic magnetic field line of the planetary magnetosphere. The reconnected field lines become open which end in the low altitude boundaries of the southern and northern hemisphere, respectively. A "reconnection site" at which reconnection takes place is surrounded by four magnetic topologies: IMF, closed, open in the south, and open in the north. A "reconnection separatrix" is then generated by connecting all the reconnection sites identified based on the field topology.

In the Jovian magnetosphere under steady-state IMF driving without considering the effect of dipole tilt, the reconnection separatrix is a continuous curve along the magnetopause, tilted from the equatorial plane. In general, the orientation of the separatrix is dependent on the IMF clock angle, and the angle of the reconnection separatrix relative to the dipole moment, i.e., the positive z-axis in the case of Earth, is half of the IMF clock angle θ defined as atan($B_z/B_y$) (*28*).

We use a hemisphere-marching method to identify the reconnection separatrix on the simulated

Jovian magnetopause (*48*).The algorithm starts from one magnetic null point and ends in the other null point on the other side of the noon-midnight meridional plane. Note that while the searching for magnetic null point is not an easy task especially under dynamic IMF conditions, we found that except for parallel (southward) IMF driving, the Jovian reconnection separatrix crosses the vicinity of the subsolar point regardless of the IMF clock angle θ.

In the separatrix-tracing algorithm, the starting point is selected on the last closed field line that intersects with the equatorial plane at 12 magnetic local time (MLT). Centered at the starting point, a dawnward-facing hemispheric surface with a radius of 5 $R_J$ is sliced into a 60×60 longitude-latitude mesh. Magnetic field lines are traced starting from these 3600 grid points to determine the distribution of magnetic topology surrounding the starting point. The one-sided hemisphere mesh is chosen such that it guarantees to sample both the magnetosphere and the magnetosheath. Based on the magnetic field topology, one block with northern hemispheric open field line connections and another block with southern hemispheric open field line connections can be identified on the hemisphere surface. The midpoint of the two points from the two blocks is regarded as a point right on the reconnection separatrix. This midpoint is used as the center of the next hemispheric surface, on which the identification process is repeated. The hemispheric surface marches along the magnetopause toward the dawnside until one or both of the open topology blocks vanishes. A duskward-facing hemispheric surface is then created for the duskward marching, which is complimentary to the first hemispheric surface in the dawnward marching.

Figure S4 shows the reconnection separatrix along the magnetopause traced using the hemisphere-marching technique. It should be pointed out that the separatrix is not in the equatorial plane since the IMF deviates from the east-west orientation. The reconnection potential is calculated by integrating the parallel component of the electric field $E_\parallel$ along the reconnection separatrix. Using the average electric and magnetic field calculated between Spin 21-23, the integral method gives approximately 508 kV as the reconnection potential on the dayside magnetopause.

*B.2 Possible influence of hot plasma populations*

The current global simulation of the Jovian magnetosphere lacks a non-thermal (hot) plasma population in the outer magnetosphere, which is unlikely described by resistive MHD. Thus, the simulated stand-off distance of the Jovian magnetosphere ($\approx$ 50 $R_J$) is lower than values from in-situ observations (> 60 $R_J$), which is determined by the balance between the SW dynamic pressure and the magnetic pressure of the Jovian dayside magnetosphere. The implementation of such a non-thermal, hot plasma population requires including non-MHD physical treatment, which is yet to be developed and has not been implemented in any of the current numerical simulations of the global Jovian magnetosphere. The underlying physical processes and numerical experiments on adding the non-thermal plasma population in the outer Jovian magnetosphere is a separate topic, which will be investigated in follow-up studies. Here we discuss briefly the possible influences of the missing hot population on the dayside SW-Magnetosphere interaction based on the balance of magnetosheath forces (*49*).

If a hot plasma population is included in the simulation, the size of the jovian magnetopause increases with a greater stand-off distance and a blunter-shaped magnetopause due to the increased total pressure in the outer jovian magnetosphere. If the width of the magnetosheath did not change with the inflated magnetosphere (*43*), the Jovi-effective length in the dayside solar wind is expected to increase proportionally as the size of the magnetosphere, resulting in an increase of approximately 15−20% in the dayside merging potential. However, since the width of the magnetosheath is also proportional to the size of the magnetosphere, the increase in the magnetosheath width diverts more solar wind flux, resulting in a decrease in the dayside merging potential (*50*). On the other hand, numerical experiments have shown that a blunter magnetosphere causes additional diversion of the solar wind flux tubes, which further suspends the merging of solar wind flux. As a consequence, the 15 − 20% enhancement in the dayside merging potential is possibly an overestimation of the effect from including a hot population. Thus, although the hot plasma population has potentially a significant effect on the size of the simulated Jovian magnetosphere, its quantitative influence on the dayside merging between the SW and the Jovian magnetosphere is possibly less than 10%. Future studies are needed to provide more accurate estimations.

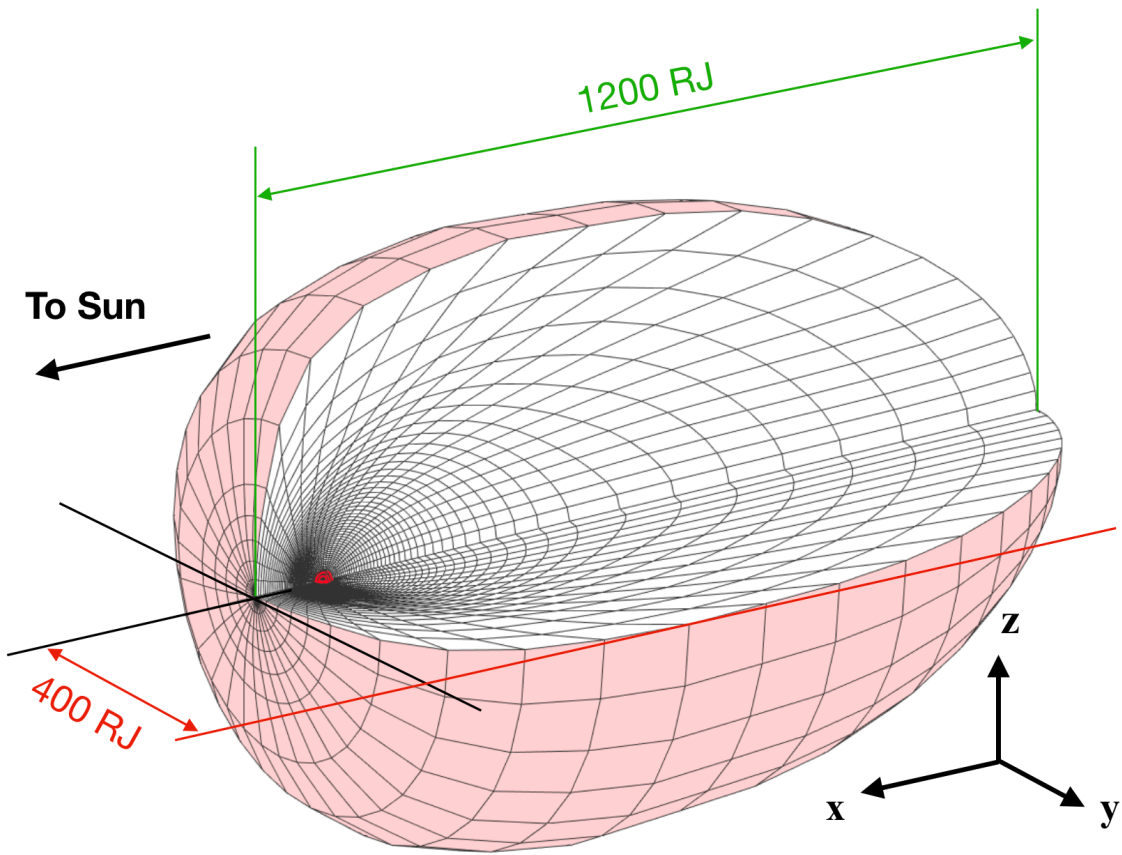

**Fig. S1.** The 3-D view of the stretched spherical grid with the distribution of the grid cells in the equatorial and meridional planes. The grid resolution showing contains 64×64×64 computation cells while the actual calculation uses 256×256×256 (radial×meridional× azimuthal) cells in the spherical coordinates.

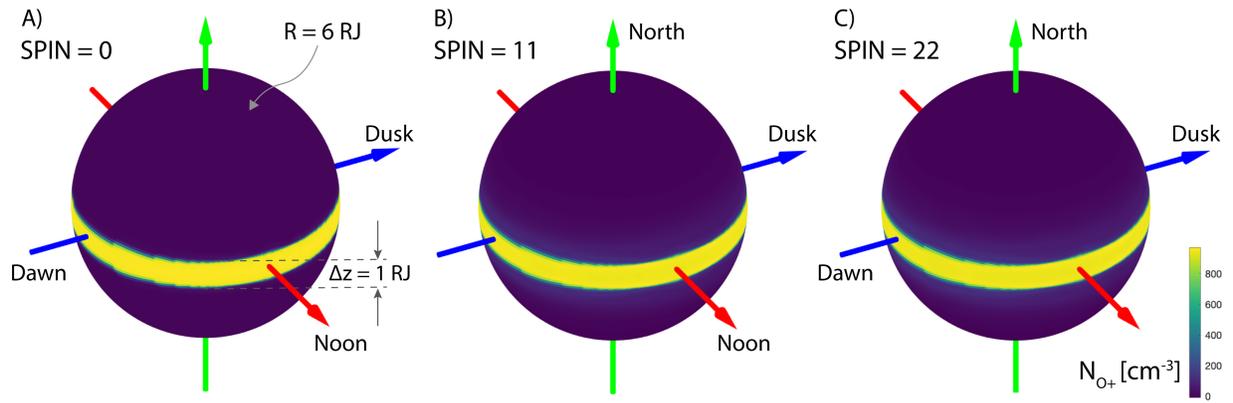

**Fig. S2.** The spatial distribution of Io mass loading introduced at the low-altitude (6 $R_J$) boundary of the simulation domain at A) t=0; B) t = 100 hour (spin 11) and C) t = 200 hour (spin 22)

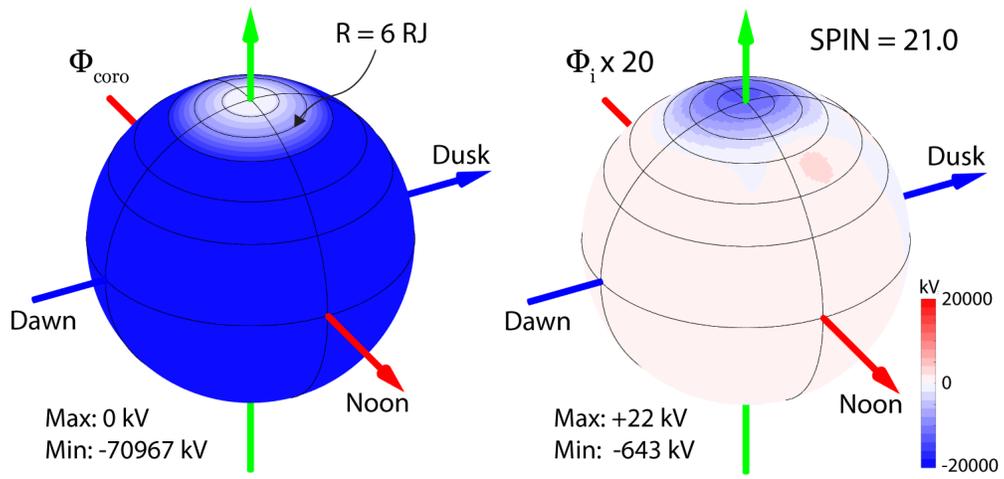

**Fig. S3.** The corotation potential (left) versus convection potential (right) at t = 203.5 hour.

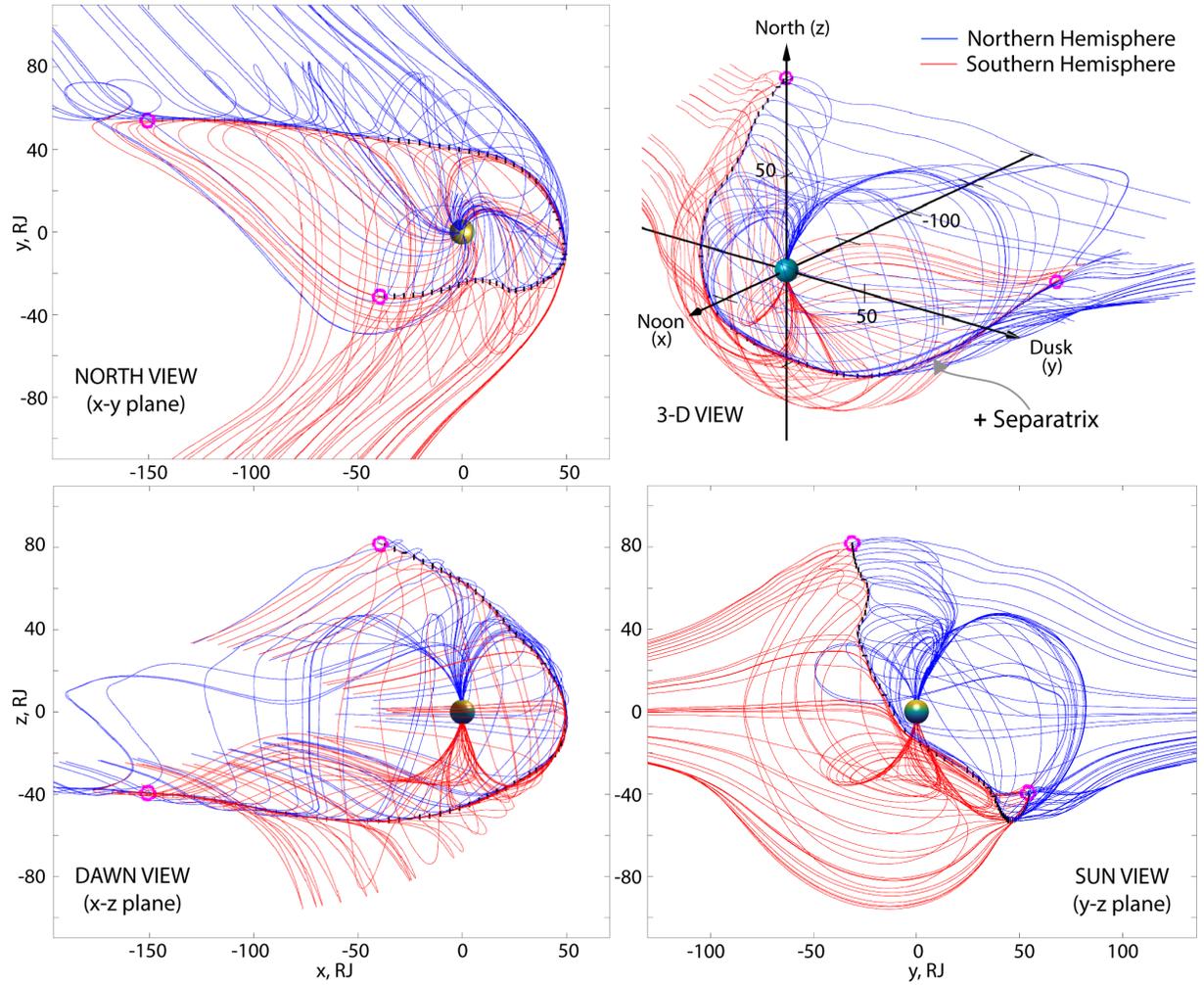

**Fig. S4.** The magnetic reconnection separatrix (black) viewing from the north (top left), the dawnside (bottom left), and the Sun (bottom right). The corresponding three-dimensional view is shown in the top right panel. The blue (red) field lines have ionospheric footprints in the northern (southern) hemisphere. The magenta points indicate the terminations of the magnetopause reconnection separatrix.